\documentclass[twocolumn,british,pra,aps,showpacs]{revtex4-1}
\usepackage[letterpaper]{geometry}
\usepackage{url}
\usepackage{graphicx}
\usepackage{amsmath}
\usepackage{amssymb}
\usepackage{bbold,soul}
\usepackage{esint}
\usepackage{epstopdf}
\usepackage{subfigure}
\usepackage{color}
\usepackage{babel}
\newcommand{\bra}[1]{\langle #1|}
\newcommand{\ket}[1]{|#1\rangle}
\newcommand{\braket}[2]{\langle #1|#2\rangle}

\newcommand{\figref}[1]{fig.~\ref{#1}}

\definecolor{nblue}{rgb}{0.3,0.3,1.0}
\definecolor{ngreen}{rgb}{0.2,0.7,0.2}
\definecolor{nred}{rgb}{0.9,0.1,0}
\definecolor{norange}{rgb}{0.8,0.5,0}

\begin{document}

\title{Classical capacity per unit cost for quantum channels}

\author{Marcin Jarzyna}\email{m.jarzyna@cent.uw.edu.pl}
\affiliation{Centre of New Technologies, University of Warsaw, Ulica Banacha 2c, 02-097 Warsaw, Poland}
\date{\today}

\begin{abstract}\noindent
In most communication scenarios, sending a symbol encoded in a quantum state requires spending resources such as energy, which can be quantified by a cost of communication. A standard approach in this context is to quantify the performance of communication protocol by classical capacity, quantifying the maximal amount of information that can be transmitted through a quantum channel per single use of the channel. However, different figures of merit are also possible, and a particularly well-suited one is the classical capacity per unit cost, which quantifies the maximal amount of information that can be transmitted per unit cost. I generalize this concept to account for the quantum nature of the information carriers and communication channels and show that if there exists a state with cost equal to zero, e.g. a vacuum state, the capacity per unit cost can be expressed by a simple formula containing maximization of the relative entropy between two quantum states. This enables me to analyze the behavior of photon information efficiency for general communication tasks and show simple bounds on the capacity per unit cost in terms of quantities familiar from quantum estimation theory. I calculate also the capacity per unit cost for general Gaussian quantum channels.
\end{abstract}

\maketitle




\section{Introduction}

In the theory of communication \cite{Shannon1948} it is common to consider how much information can be transmitted from sender to receiver per single channel use, which is quantified by the information transmission rate. In that picture many fundamental results have been obtained concerning both specific transmission protocols \cite{Chen2012, Chung2011, Guha2011, Guha2011a, Takeoka2014, Wilde2012} and general optimal bounds \cite{Shapiro2004, Palma2014,Giovannetti2004a, Giovannetti2014, Lupo2010, Shapiro2004, Shapiro2009}. The application of laws of quantum mechanics has allowed for a deeper analysis of communication \cite{Gordon1962, Holevo1973, Hausladen1996, Schumacher1997} and showing such effects as output and input superadditivity \cite{Sasaki1998, Hastings2009, Czajkowski2016} or finite optimal rates even for noiseless channels, emerging from nonclassical phenomena such as entanglement or the Heisenberg uncertainty principle.

Most of the work mentioned above has been devoted to understanding quantum effects affecting information transmission rates and how they can enhance the capacity, i.e. the maximal amount of bits transmitted per single channel use. There are, however, alternative figures of merit that quantify the efficiency of communication protocols not only in relation to the number of channel uses but also when some other physical restrictions are taken into account. Indeed, in most communication schemes there are usually some constraints on input symbols or features of information carriers that can be used during the information transmission task. These are usually quantified by the cost of using a particular state that encodes transmitted symbol. A particularly important and often used example of such a cost is the average energy of the signal. In this context the quantity of interest is usually a capacity-cost function, quantifying the maximum number of bits that can be transmitted per channel use with an average cost not exceeding some given value. However, instead of considering how many bits can be transmitted by a single channel use, it may be more informative to express the performance of the protocol by the efficiency, which is the maximal number of bits that can be transmitted per unit cost. Such a quantity is called the capacity per unit cost \cite{Verdu1990} and is especially interesting in many instances in which it is the cost of sending a symbol crucial limitation rather than the number of channel uses. 

A basic example of situation in which the capacity per unit cost is particularly meaningful is long-distance or even deep-space communication, important for space mission design \cite{Hemmati2011, Waseda2011, Guha2011a, Chen2012, Powell2013}. Since it is the energy budget that is the most sensitive resource in such settings it is crucial to increase the amount of information transmitted with a single unit of energy. Another example of a situation in which capacity per unit cost is well suited for the description of communication tasks is when the energy is spend not only on the modulation procedure, that is, generating a signal, but also on the propagation of the information carriers like in various signal regeneration techniques \cite{Yariv1990, Antonelli2014, Sorokina2014}. In such instances it is crucial to take into account the energy spent on regeneration, since it may modify the overall performance of the protocol.

In this work I generalize the concept of capacity per unit cost, discussed in a classical setting in \cite{Verdu1990}, to account for the quantum nature of the information transmission task. In particular I give a simple formula for the capacity per unit cost for any quantum channel assuming the existence of a cost-free state. This is a common situation in actual communication systems in which the cost is usually the energy of the light pulse and one can use an empty vacuum pulse for free. In such a case I show that the  capacity per unit cost can be bounded by a quantum Fisher information, a quantity well known in quantum estimation theory. Finally, I give also a detailed analysis of the capacity per unit cost for general Gaussian quantum channels.

\begin{figure}[t!]
\includegraphics[width=\columnwidth]{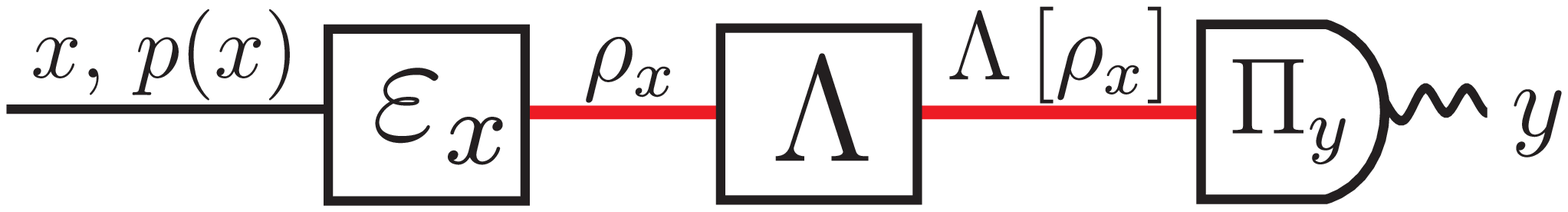}
\caption{Basic scheme of classical communication through a quantum channel. Input symbols $x$, distributed according to the prior probability distribution $p(x)$, are encoded into quantum states $\rho_x$ of information carriers by a modulation operation, represented by a quantum channel $\varepsilon_x$. The noisy communication link is modeled by some quantum channel $\Lambda$, which outputs states $\Lambda[\rho_x]$, which are then measured by a positive-operator-valued measure $\Pi_y$, resulting in an output symbol $y$ used to reconstruct the input message.}
\label{fig:scheme}
\end{figure}

\section{Communication theory}

Mathematical foundations of communication theory were defined by Shannon in a seminal paper \cite{Shannon1948}. The main goal in communication is to faithfully transmit as much information as possible from the sender to receiver. This is a nontrivial problem since usually environmental noise and other experimental imperfections corrupt the signal which makes it hard to decode. A basic information transmission task is schematically visualized in \figref{fig:scheme}. A message is encoded into a string of symbols $x$ distributed according to a probability distribution $p(x)$, describing the encoding. Every symbol is then encoded in a quantum state $\rho_x$ of an information carrier by some modulation operation that can be described by a quantum channel $\varepsilon_x$. Information carriers are transmitted by a noisy communication link and undergo a decoherence process modeled by a quantum channel $\Lambda$. The output state $\Lambda[\rho_x]$ is then measured which is represented by a general positive-operator-valued measure $\Pi_y$. Finally, the results of the measurement $y$ are used to decode the initial message. The performance of such a scheme for a given code described by $p(x)$ is given by mutual information
\begin{equation}\label{eq:mutual_information}
I(X,Y)=H(Y)-H(Y|X),
\end{equation}
where $H(Y)=-\sum_y p(y)\log p(y)$ is the Shannon entropy of the output distribution and $H(Y|X)=-\sum_{x,y}p(x)p(y|x)\log p(y|x)$ is the conditional entropy. Note that mutual information depends on the actual measurement and set of states used for transmission through the conditional probability distribution $p(y|x)=\textrm{Tr}(\Lambda[\rho_x]\Pi_y)$.  Optimization of the above expression over different encodings (that is, prior probability distributions) gives the best possible rate for a given set of quantum states $\{\rho_x\}$ and measurement $\Pi_y$, known as capacity
\begin{equation}\label{capacity_classical}
C=\sup_{p(x)}I(X,Y).
\end{equation}
Importantly, mutual information and capacity give the amount of information that can be transmitted {\it perfectly} per single use of a noisy channel, i.e., without any disturbance. Allowing for non negligible level of errors in the decoded message can increase the information transmission rate, which is described by rate distortion theory \cite{Shannon1959, Cover1991, Nair2012}.

The quantum information picture allows one to optimize not only over encodings, but also detection apparatus and states used in the protocol. Quantum measurements may be collective, i.e., performed on the outputs of a large number of channel uses, which in principle can enhance the communication performance over the best single-shot strategy, the phenomenon known as output superadditivity \cite{Sasaki1998, Shor2002, Czajkowski2016}. Therefore, a valid quantity describing communication performance is a regularized mutual information optimized over (possibly collective) output measurements
\begin{equation}
I_{\textrm{ac}}(X^n,Y^n)=\sup_{\{\Pi_{y^n}\}}\frac{I(X^n,Y^n)}{n},
\end{equation}
which is known as accessible information. Going to a large number of channel uses $n\to\infty$ and using arbitrarily large collective measurements allows for communication at the rate $\chi=\lim_{n\to\infty}I_{\textrm{ac}}(X^n,Y^n)$, known as Holevo information \cite{Holevo1973} and given by
\begin{equation}\label{eq:holevo1}
\chi=S\left(\sum_x p(x)\Lambda[\rho_x]\right)-\sum_x p(x)S(\Lambda[\rho_x]),
\end{equation}
where $S(\rho)=-\textrm{Tr}\rho\log\rho$ denotes the von Neumann entropy of state $\rho$. For the purpose of this article it is convenient to rephrase the expression for Holevo information in terms of the quantum relative entropy
\begin{equation}\label{eq:holevo2}
\chi=\sum_x D\left(\Lambda[\rho_x]||\Lambda[\bar{\rho}]\right)p(x),
\end{equation}
where $D(\rho||\sigma)=\textrm{Tr}(\rho\log\rho-\rho\log\sigma)$ is the quantum relative entropy between states $\rho$ and $\sigma$. The expressions (\ref{eq:holevo1}) and (\ref{eq:holevo2}) can be easily generalized to the continuous encodings by replacing the sum with an integral.

Analogously, it is in principle possible to increase the communication rate by allowing for entangled input sates, which is known as the input superadditivity \cite{Hastings2009}. Eventually, the ultimate rate at which  it is possible to faithfully communicate information is given by a regularized expression
\begin{equation}
C=\lim_{n\to\infty}\sup_{p(x^n),\rho_{x^n}}\frac{\chi\left(\{\Lambda^{\otimes n}[\rho_{x^n}]\}\right)}{n},
\end{equation}
which is known as the classical capacity of a channel $\Lambda$ and where the supremum is taken over the whole set of entangled states between $n$ inputs. Crucially, in contrast to classical scenarios, in the quantum picture the optimal measurements and input states (which may both be entangled) provide a conditional probability distribution that is not separable, $p(y^n|x^n)\neq p(y_1|x_1)\dots p(y_n|x_n)$. This makes analysis difficult as various classical limit theorems do not apply.

\section{Capacity per unit cost}

In most communication schemes there are usually constraints on the total amount of resources that can be utilized in the protocol. The easiest way to model such constraints is through the introduction of the cost of a given symbol $x$, quantified by the cost function $b[x]$, which associates a nonnegative cost with every value of $x$. Instead of capacity, which quantifies how many natural units of information can be transmitted per channel use, in such cases the quantity of interest is rather the capacity-cost function $C(\beta)$, describing the maximum number of natural units of information per channel use that can be transmitted with an average cost not exceeding some given value $\beta$. In such a picture the transmission of one natural unit of information requires $1/C(\beta)$ symbols at the average cost $\beta/C(\beta)$. In terms of efficiency an important quantity in this context is the minimum cost required for the transmission of a single natural unit of information of information through a channel or its reciprocal, the capacity per unit cost, which I will denote by $\mathbb{\mathbf{C}}$. In \cite{Verdu1990} it was shown how to calculate $\mathbf{C}$ in many important cases in classical information theory. Here I will show how this concept can be adapted to the underlying quantum description of information carriers and measurement devices. Many definitions that I invoke below are taken directly from \cite{Verdu1990} in order to state the results.

An $(n,M,\nu,\epsilon)$ is a code with block length $n$, the number of all codewords equal to $M$, and the cost of each codeword $(x_{m,1},\dots x_{m,n})$ bounded from above by $\nu$, i.e.,
\begin{equation}
\sum_{i=1}^{n}b[x_{m,i}]\leq\nu
\end{equation}
The probability of correct decoding of any message in the code is greater than $1-\epsilon$. With such a definition it is possible to state what is the rate with cost per symbol not exceeding $\beta$. For $0\leq\epsilon\leq1$ and $\beta>0$ we call $R>0$ an $\epsilon$-achievable rate with cost per symbol not exceeding $\beta$ if for every $\gamma>0$ there exist $n_{0}$ such that for every $n>n_{0}$ one can find a code $(n,M,n\beta,\epsilon)$ for which $\log M>n(R-\gamma)$. If $R$ is $\epsilon$-achievable for every $\epsilon$ it is called just an achievable rate and the maximal such rate is the capacity with cost per symbol not exceeding $\beta$, quantified by the capacity-cost function \textbf{$C(\beta)$}. For a memoryless stationary channel, the capacity-cost function can be easily stated as 
\begin{equation}\label{eq:capacity_cost}
C(\beta)=\sup_{p(x),\langle b[x]\rangle\leq\beta}I(X,Y),
\end{equation}
where $\langle b[x]\rangle=\sum_x p(x)b[x]$ is the average cost. In other words, $C(\beta)$ is just the maximal mutual information under constraint of a given maximal input's average cost. Since increasing the maximal allowed value of the average cost $\beta$ cannot decrease the information transmission rate, the capacity-cost function is nondecreasing. Similarly, it may also be shown to be concave \cite{Verdu1990}. As noted in the preceding section, in the quantum mechanical description a conditional probability distribution arises through the Born rule $p(y|x)=\textrm{Tr}(\Lambda[\rho_x]\Pi_y)$ so it depends on the input state $\rho_x$ and output measurement $\Pi_y$. It is therefore possible to transform eq.~(\ref{eq:capacity_cost}) further, since optimization can be performed also over quantum states at the input and measurements 
\begin{multline}\label{eq:ref1}
C^{(n)}(\beta)=\sup_{\{ p(x^n),\rho_{x^{n}},\Pi_{y^{n}}\},\langle b[\rho_{x^n}]/n\rangle\leq\beta}\frac{I(X^{n},Y^{n})}{n}=\\
=\sup_{\left\{ p(x^{n}),\rho_{x^{n}}\right\} ,\langle b[\rho_{x^n}]/n\rangle\leq\beta}I_{\textrm{ac}}(X^{n},Y^{n})\leq\\
\leq\sup_{\{p(x^{n}),\rho_{x^{n}}\},\langle b[\rho_x^n]/n\rangle\leq\beta}\frac{\chi\left[\{\Lambda^{\otimes n}[\rho_{x^{n}}]\}\right]}{n},
\end{multline}
where the index $n$ explicitly denotes superadditivity of (regularized) capacity $C^{(n)}$, accessible information $I_{\textrm{ac}}(X^{n},Y^{n})$ and Holevo information and $b[\rho]$ denotes the cost of using state $\rho$. Note that the last inequality in eq.~(\ref{eq:ref1}) is saturable in the limit of long messages by performing an appropriate collective measurement. In order to get the ultimate capacity-cost function I will take the limit of long messages
\begin{equation}
C(\beta)=\lim_{n\to\infty}C^{(n)}(\beta). 
\end{equation}

For $0\leq\epsilon\leq1$ I will call $\mathbf{R}>0$ an $\epsilon$-achievable rate per unit cost if for every $\gamma>0$ there exist $\nu_{0}\in\mathbb{R}_{+}$ such that for every $\nu>\nu_{0}$ one can find a code $(n,M,\nu,\epsilon)$ for which $\log M>\nu(R-\gamma)$. If $R$ is $\epsilon$-achievable for every $\epsilon$ it is called just an achievable rate and maximal such rate is a capacity per unit cost. One may show that $\mathbf{C}$ is given by
\begin{equation}\label{eq:capacity_quantum}
\mathbf{C}=\sup_{\beta>0}\frac{C(\beta)}{\beta}=\lim_{n\to\infty}\sup_{\{p(x^{n}),\rho_{x^{n}}\}}\frac{\chi\left[\Lambda^{\otimes n}[\rho_{x^{n}}]\right]}{\langle b[\rho_{x^n}]\rangle}.
\end{equation}
Note that normalization to the number of channel uses in the last expression is included in the fact that I consider the cost of using the input state for many channel uses $\rho_{x^n}$. The proof is similar to the one considering the classical case in \citep{Verdu1990} with the exception that in the present case the classical channel transforming the input alphabet to the output one is not memoryless. This is because one can use entangled states and make collective measurement so the joint probability does not factorize into the product of probabilities  in each channel use. However, I will show that with the assumption that the quantum channel $\Lambda$ transforming input states to the output ones is memoryless, i.e., it acts in the same way irrespective of what the previous input was, eq.~(\ref{eq:capacity_quantum}) still applies. In \cite{Verdu1990} the proof of the classical result was divided into two steps, one in which it was shown that $C(\beta)/\beta$ is an achievable rate and the converse part. Since the assumption of memorylessness was crucial only to the second part of this proof, the first step remains unchanged and I will present here only the second step. 

By the Fano inequality \cite{Fano1961} one can write that $(1-\epsilon)\log M\leq I(\tilde{X}^{n},\tilde{Y}^{n})+\log2$, where $\tilde{X}^{n}$ is the distribution of the input symbols when messages are equiprobable. Since $\tilde{X}^{n}$ satisfies the condition for cost $\langle b[\rho_{\tilde{x}^n}]\rangle\leq\nu$ it is possible to write the chain of inequalities 
\begin{multline}\label{eq:ref2}
\frac{\log M}{\nu}=\frac{1}{1-\epsilon}\left[\frac{n}{\nu}\frac{I(\tilde{X}^{n},\tilde{Y}^{n})}{n}+\frac{\log2}{\nu}\right]\leq\\
\leq\frac{1}{1-\epsilon}\left[\frac{n}{\nu}\sup_{p(x^{n}),\,\langle b[\rho_{x}^n]/n\rangle\leq\frac{\nu}{n}}\frac{I(X^{n},Y^{n})}{n}+\frac{\log2}{\nu}\right]\leq\\
\leq\frac{1}{1-\epsilon}\left[\frac{n}{\nu}\sup_{\substack{\left\{ p(x^{n}),\rho_{x^{n}}\right\},\\ \langle b[\rho_{x}^n]/n\rangle\leq\frac{\nu}{n}}}\frac{\chi\left[\Lambda^{\otimes n}\left[\rho_{x^{n}}\right]\right]}{n}+\frac{\log2}{\nu}\right]\leq\\
\leq\frac{1}{1-\epsilon}\left[\sup_{\beta>0}\frac{C(\beta)}{\beta}+\frac{\log2}{\nu}\right],
\end{multline}
where the second inequality comes from the fact that accessible information is upper bounded by the Holevo information and in the last one I have used $C^{(n)}(\beta)\leq C(\beta)$. As a consequence of the inequality (\ref{eq:ref2}) for any $\epsilon$-achievable rate $\mathbf{R}$ for every $\gamma>0$ there exist $\nu_{0}$ such that for $\nu>\nu_{0}$
\begin{equation}
\mathbf{R}-\gamma<\frac{1}{1-\epsilon}\left[\sup_{\beta>0}\frac{C(\beta)}{\beta}+\frac{\log2}{\nu}\right].
\end{equation}
Since $\nu$ in the above expression can be arbitrarily large, the rate $\mathbf{R}$ is achievable if $\mathbf{R}\leq\sup_{\beta>0}\frac{C(\beta)}{\beta}$. The proof in the other direction does not require a memorylessness assumption and is the same as in the classical case \cite{Verdu1990}, so I will not present it here.

Importantly, note that one can redo the above calculations with an additional restriction to only some class of input states. In particular, for the case of separable input states  $\rho_{x^{n}}=\otimes_{i}\rho_{x_{i}}$ this means that one may write
\begin{equation}\label{eq:capacity_separable}
\mathbf{C}_{\textrm{sep}}=\sup_{\beta>0}\frac{C_{\textrm{sep}}(\beta)}{\beta}=\sup_{\left\{p(x),\rho_{x}\right\} }\frac{\chi\left[\Lambda\left[\rho_{x}\right]\right]}{\langle b[\rho_x]\rangle}.
\end{equation}
Therefore, since most realistic communication protocols use separable states in the rest of the paper I will focus on consequences of eq.~(\ref{eq:capacity_separable}). The discussion, however, will be applicable also to full capacity. To obtain results in that case it will be enough to take regularized expressions everywhere. 
 
A particularly important instance in which eq.~(\ref{eq:capacity_separable}) can be applied is when there is a free state $\rho_0$ in the input ensemble. In such a case eq.~(\ref{eq:capacity_separable}) can be greatly simplified and written as
\begin{equation}\label{eq:cap_relative}
\mathbf{C}=\sup_{\rho_x\neq\rho_0}\frac{D(\Lambda[\rho_{x}]||\Lambda[\rho_{0}])}{b[\rho_x]}.
\end{equation}
To show this, I will once again refer to the proof of the classical result made in \cite{Verdu1990} by adapting it to the quantum framework. First let me note that since $C(\beta)$ is concave on the interval $(0,\infty)$, the function $C(\beta)/\beta$ is nonincreasing on this interval. This means that we can omit the supremum in eq.~(\ref{eq:capacity_quantum}) and write
\begin{equation}\label{eq:cap_beta_0}
\mathbf{C}=\lim_{\beta\to0}\frac{C(\beta)}{\beta}.
\end{equation}
Let me now consider two cases:
\vskip.2cm
1. If there is more than one free input state, then it is possible to encode all the information for free using only these states, which means $\mathbf{C}=\infty$. On the other hand, as long as output states are distinguishable  $\Lambda[\rho_{x}]\neq\Lambda[\rho_{0}]$, the relative entropy is strictly positive $D(\Lambda[\rho_{x}]||\Lambda[\rho_{0}])>0$, which means that $\sup_{\rho_x\neq\rho_0}D(\Lambda[\rho_{x}]||\Lambda[\rho_{0}])/b[\rho_x]=\infty$.
\vskip.2cm
2. If there is only one free input symbol the proof is more complicated. First, let me calculate the Holevo information achieved by the binary input distribution $p(1)=\frac{\beta}{b[\rho]}=1-p(0)$, for arbitrary state $\rho\neq\rho_0$
\begin{multline}
\chi=\int D(\Lambda[\rho_{x}]||\Lambda[\bar{\rho}])p(x)dx=\\
=\int dxp(x)\textrm{Tr}\left[\Lambda[\rho_{x}]\log\Lambda[\rho_{x}]-\Lambda[\rho_{x}]\log\Lambda[\bar{\rho}]\right]=\\
=\int dxp(x)\textrm{Tr}\left[\Lambda[\rho_{x}]\log\Lambda[\rho_{x}]-\Lambda[\rho_{x}]\log\Lambda[\rho_{0}]\right]+\\
-\int dxp(x)\textrm{Tr}\left[\Lambda[\rho_{x}]\log\Lambda[\bar{\rho}]-\Lambda[\rho_{x}]\log\Lambda[\rho_{0}]\right]=\\
=\int dxp(x)D\left(\Lambda[\rho_{x}]||\Lambda[\rho_{0}]\right)-D\left(\Lambda[\bar{\rho}]||\Lambda[\rho_{0}]\right),\label{eq:holveo_long}
\end{multline}
where $\bar{\rho}=\int dxp(x)\rho_{x}$ is the averaged state. For the particular encoding that I choose this means that
\begin{equation}\label{eq:cap_entr_proof1}
\frac{\chi}{\beta}=\frac{1}{b[\rho]}D\left(\Lambda[\rho]||\Lambda[\rho_{0}]\right)-\frac{1}{\beta}D\left(\Lambda[\bar{\rho}]||\Lambda[\rho_{0}]\right).
\end{equation}
The only part depending on $\beta$ on the right-hand side of the above equation is the second term which can  be rewritten as $\frac{1}{\beta}D\left(\frac{\beta}{b[\rho]}\Lambda[\rho]+\left(1-\frac{\beta}{b[\rho]}\right)\Lambda[\rho_{0}]||\Lambda[\rho_{0}]\right)$. For small $\beta$ in the leading order it is given by
\begin{multline}\label{eq:Cap_entropy_proof_J}
D\left(\frac{\beta}{b[\rho]}\Lambda[\rho]+\left(1-\frac{\beta}{b[\rho]}\right)\Lambda[\rho_{0}]||\Lambda[\rho_{0}]\right)=\\
=\frac{\beta^{2}}{2b[\rho]^{2}}\mathcal{J}_{\theta=0}\left[\rho_\theta]\right].
\end{multline}
The quantity $\mathcal{J}_{\theta=0}\left[\rho_\theta\right]$ appearing in the above formula is a relative entropy quantum Fisher information (REQFI) \cite{Petz1996, Koenig2013, Czajkowski2016} at $\theta=0$ for a continuously parametrized family of quantum states $\rho_\theta=(1-\theta)\Lambda[\rho_{0}]+\theta\Lambda[\rho]$. In general, for an arbitrary family of states $\rho_\varphi$ parametrized by a continuous parameter $\varphi$ REQFI can be read out from the second-order term in the Taylor expansion of the relative entropy between states $\rho_\varphi$ and $\rho_{\varphi+\delta\varphi}$, i.e. $D(\rho_\varphi||\rho_{\varphi+\delta\varphi})\approx \frac{\delta\varphi^2}{2}\mathcal{J}[\rho_\varphi]$. I give an explicit expression for REQFI $\mathcal{J}[\rho_\varphi]$ of an arbitrary parametrized family of states $\rho_\varphi$ in the Appendix \ref{app:a}. If $\Lambda[\rho_{0}]$ and $\Lambda[\rho]$ have the same support (i.e. when $D\left(\Lambda[\rho]||\Lambda[\rho_{0}]\right)$ is finite) then $\mathcal{J}_{\theta=0}[\rho_\theta]$ is finite. Thus, since according to eq.~(\ref{eq:cap_beta_0}) capacity is attained in the limit $\beta\to 0$, I can write
\begin{equation}\label{eq:cap_proof_direct}
\mathbf{C}=\lim_{\beta\to0}\frac{\chi}{\beta}\geq\sup_{\rho\neq\rho_0}\frac{1}{b[\rho]}D\left(\Lambda[\rho]||\Lambda[\rho_{0}]\right).
\end{equation}
To prove the equality I will consider the inequality $\chi\leq\int dxp(x)D\left(\Lambda[\rho_{x}]||\Lambda[\rho_{0}]\right)$, which follows from eq.~(\ref{eq:holveo_long}) and the fact that the relative entropy is a nonnegative quantity
\begin{multline}\label{eq:cap_proof_inverse}
\frac{C(\beta)}{\beta}=\frac{1}{\beta}\sup_{\{\rho_x\},\langle b[\rho_x]\rangle\leq\beta}\chi\leq\\
\leq\sup_{\{\rho_x\},\langle b[\rho_x]\rangle\leq\beta}\int dxp(x)\frac{D\left(\Lambda[\rho_{x}]||\Lambda[\rho_{0}]\right)}{b[\rho_x]}\frac{b[\rho_x]}{\beta}\leq\\
\leq\sup_{\rho\neq\rho_0}\frac{D\left(\Lambda[\rho]||\Lambda[\rho_{0}]\right)}{b[\rho]},
\end{multline}
where I have used the fact that $\frac{\int p(x)b[\rho_x]dx}{\beta}\leq1$. Equations~(\ref{eq:cap_proof_direct}) and (\ref{eq:cap_proof_inverse}) imply eq.~(\ref{eq:cap_relative}).

Finally, to end the proof it is sufficient to see what happens if there is only a single free symbol but output states $\Lambda[\rho_0]$ and $\Lambda[\rho]$ in the above encoding have different, i.e. not fully overlapping, supports. In such a case the relative entropy between $D(\Lambda[\rho]||\Lambda[\rho_0])$ and REQFI in eq.~(\ref{eq:Cap_entropy_proof_J}) is infinite and the right hand side of eq.~(\ref{eq:cap_entr_proof1}) is not well defined. According to the first line in eq.~(\ref{eq:holveo_long}), for the encoding described above it is possible to write 
\begin{equation}
\frac{\chi}{\beta}\geq \frac{D(\Lambda[\rho]||\Lambda[\bar{\rho}])}{b[\rho]}.
\end{equation}
Since for $\beta\to 0$ the averaged state converges to the free symbol state, the right-hand side of the above equation goes to infinity and consequently also the capacity per unit cost. Thus eq.~(\ref{eq:cap_relative}) applies also in this case. This ends the proof.
\vskip.2cm

A crucial observation from eq.~(\ref{eq:cap_relative}) is that if $\Lambda[\rho_{0}]$ is pure then, assuming the channel does not act trivially, i.e., there exist a state $\rho$ such that $\Lambda[\rho]\neq\Lambda[\rho_0]$, the channel capacity per unit cost $\mathbf{C}$ is always infinite. This is because the relative entropy between two states with partially overlapping supports is infinite, as discussed above. In most realistic protocols $b[\rho_x]$ is the energy, or average number of photons per symbol $\bar{n}$ in the state $\rho_x$ used for communication and therefore there usually exists a free symbol that costs no energy: a vacuum state $\rho_{0}=|0\rangle\langle0|$. In this context a quantity closely related to capacity per unit cost is the photon information efficiency (PIE) defined as $\Pi(\bar{n})=C(\bar{n})/\bar{n}$. If the channel $\Lambda$ does not add any energy to the signal (for example it is a lossy channel or dephasing channel) then it does not corrupt the vacuum state $\Lambda[\rho_{0}]=|0\rangle\langle0|$. In such cases, according to eq.~(\ref{eq:cap_relative}), the capacity per unit cost is infinite, which means also that PIE is unbounded. This indicates that in a regime of small $\bar{n}$ the capacity per single use of the channel $C(\bar{n})$ has a better than linear scaling with $\bar{n}$, possibly log-linear scaling $\sim\bar{n}\log\frac{1}{\bar{n}}$ \cite{Jarzyna2014, Jarzyna2015, Chung2016, Jarzyna2016, Czajkowski2016, Rosati2016}. On the other hand, if the channel corrupts the vacuum state so that $\Lambda[|0\rangle\langle 0|]$ has the same support as other output states then the relative entropy, and consequently also the capacity per unit cost and PIE, is finite. In such a case, the capacity per single use of the channel cannot scale better than linearly for small average numbers of photons $C(\bar{n})\sim\bar{n}$. 

From the proof of eq.~(\ref{eq:cap_relative}) it is possible to conclude that if there is a free symbol in the input alphabet then the capacity per unit cost can be attained by using only two symbols, i.e., binary encoding. However, this does not mean that binary encoding is the best from the point of view of the standard capacity per single use of the channel or PIE \cite{Jarzyna2015, Chung2016}. Indeed, in the case in which $\mathbb{\mathbf{C}}=\infty$ there may exist another more complicated encoding for which PIE will be also unbounded in the limit of small cost but at the same time it will attain a higher value than for the binary encoding discussed above for any finite value of the average cost \cite{Chung2016}. This is because $\mathbb{\mathbf{C}}$ is concerned only with the maximum of PIE and not its other values. Interestingly, however, in this case it is easy to show that collective measurements cannot increase $\mathbb{\mathbf{C}}$. This is because if the free state $\rho_0$ and the second state $\rho$ used for communication have partially overlapping supports it is sufficient to use a single-shot measurement in the form of a projection onto the support of one of the states. In such a way, the conditional probability distributions for the output results $p(y|0)$ and $p(y|1)$ will have partially overlapping supports and one can easily show that the capacity per unit cost for such a classical channel is infinite \cite{Verdu1990}.

\section{Relationship with estimation theory}

An important implication of eq.~(\ref{eq:cap_relative}) is a connection that can be made between quantum estimation and communication theories \cite{Czajkowski2016}. Assuming that input symbols are encoded by a continuous parameter $x$ and the cost function is quadratic $b[\rho_x]=x^2$, the capacity per unit cost can be related to both REQFI and ordinary quantum Fisher information (QFI) \cite{Helstrom1976, Holevo1982, Braunstein1994} of the state at the output. The latter quantity lies at the heart of quantum estimation theory and for a parametrized family of states $\rho_x$ is given by $\mathcal{F}[\rho_x]=\textrm{Tr}(\rho_x L_x^2)$, where the operator $L_x$ is given implicitly by the equation $\dot{\rho}_x=\frac{1}{2}(\rho_x L_x+L_x\rho_x)$, where the overdot denotes differentiation with respect to $x$. Crucially, QFI through the quantum Cramer-Rao bound \cite{Helstrom1976, Holevo1982, Braunstein1994} gives the saturable lower bound on the mean-square error of estimation of $x$ by any unbiased estimator $\Delta x^2\geq 1/\mathcal{F}$. 

For the quadratic cost function $b[\rho_x]=x^2$ the ratio of relative entropy $D(\Lambda[\rho_x]||\Lambda[\rho_0])$ to the cost of using symbol $x$ converges to REQFI  $\lim_{x\to0}\frac{D(\Lambda[\rho_{x}]||\Lambda[\rho_{0}])}{b[\rho_x]}=\frac{1}{2}\mathcal{J}\left(\Lambda[\rho_{x}]\right)|_{x=0}$. Since according to eq.~(\ref{eq:cap_relative}) the capacity per unit cost is given by the supremum of this ratio, it is lower bounded by 
\begin{equation}\label{eq:Cap_QFI}
\mathbb{\mathbf{C}}\geq\frac{1}{2}\mathcal{J}\left(\Lambda\left[\rho_{x}\right]\right)|_{x=0}\geq\frac{1}{2}\mathcal{F}\left(\Lambda\left[\rho_{x}\right]\right)|_{x=0}
\end{equation}
where $\mathcal{F}_{x=0}$ is QFI evaluated at $x=0$. The second inequality is a consequence of the fact that $\mathcal{F}$ is the smallest type of quantum Fisher information \cite{Petz1996, Lesniewski1999} so it has to be lower than the respective REQFI. 
The above inequality is in agreement with the results of \cite{Czajkowski2016}, in which it was shown that for narrow continuous prior distributions, that is for distributions with small variance $\beta=\langle x^2\rangle\ll 1$, in the protocol utilizing states with rank $r$ of the Hilbert space with dimension $d$, the capacity can be expressed as 
\begin{equation}
C\approx \frac{\beta}{2}\mathcal{\underline{J}}-\sum_{n=r+1}^d\frac{\beta \mathcal{F}_n}{4}\log\frac{\beta \mathcal{F}_n}{4e}.
\end{equation}
where $\mathcal{\underline{J}}$ and $\mathcal{F}_n$ are very closely related to REQFI $\mathcal{J}$ and  QFI $\mathcal{F}$ respectively (see \cite{Czajkowski2016} for exact definitions). Note that for full rank output states $d=r$ and $\mathcal{\underline{J}}=\mathcal{J}$ and the second term in the above equation vanishes which means that
\begin{equation}\label{eq:cap_REQFI_eq}
C\approx \frac{\beta}{2}\mathcal{J}\overset{\beta\to 0}{\Longrightarrow} \mathbb{\mathbf{C}}=\frac{\mathcal{J}}{2}.
\end{equation}
and the leftmost inequality in eq.~(\ref{eq:Cap_QFI}) is saturated.

Finally, let me remark that if the communication protocol utilizes coherent states with amplitude $\alpha$ then the quadratic cost function $b[\ket{\alpha}]=|\alpha|^{2}$ appears naturally and is exactly the energy of the signal. Since $\mathbb{\mathbf{C}}$ is the capacity per unit cost and the cost is just the energy of the signal, the reciprocal of the capacity $1/\mathbb{\mathbf{C}}$ is equal to the minimal energy required for the transmission of a single natural unit of information. Therefore eq.~(\ref{eq:Cap_QFI}) states that the minimal energy $E_{\textrm{min}}$ required for transmission of a half nat of information is upper bounded by the precision $\Delta \alpha^2$ of the estimation of the amplitude of output signal
\begin{equation}\label{eq:min_energy}
E_{\textrm{min}}\leq\frac{1}{\mathcal{J}}\leq\frac{1}{\mathcal{F}}\leq\Delta x^2.
\end{equation}
This expression means that if it is possible to estimate the input signal with large precision then it is not necessary to use a great deal of energy at the input, which agrees with common-sense intuition. Note, however, that unlike an analogical classical expression \cite{Verdu1990}, the inequalities in eq.~(\ref{eq:min_energy}) in general are not saturable. This is because in the classical picture there exists only a single Fisher information \cite{Chentsov1978, Petz1996, Lesniewski1999}, which gives a saturable bound on precision, but in the quantum picture REQFI is usually strictly larger than QFI $\mathcal{F}\leq\mathcal{J}$ and therefore precision gives only a loose upper bound on $E_{\textrm{min}}$ .

\section{Classical capacity per unit cost for Gaussian channels}
\label{sec:gaussian}

As an example I will investigate the classical capacity per unit cost for the most general Gaussian evolution and input Gaussian states. As the cost of sending state $\rho$ I will use its energy, which I will quantify by the average number of photons in the state. In such a protocol there always exists a free symbol, which is just the vacuum state, which evolves under the Gaussian channel $\Lambda$ into $\rho_{0}=\Lambda[|0\rangle\langle0|]$. To find the capacity per unit cost I will first calculate the relative entropy between $\rho_{0}$ and $\rho_{\textrm{out}}=\Lambda[\rho]$, where $\rho$ is some single mode Gaussian state with an average number of photons $\bar{n}$.  Any such state can  be written as
\begin{equation}
\rho=D(\alpha)S(r)\rho_{\textrm{th}}S^{\dagger}(r)D^{\dagger}(\alpha),
\end{equation}
where $\rho_{\textrm{th}}=\sum_{n=0}^{\infty}\frac{\bar{N}_{\textrm{in}}^{n}}{(\bar{N}_{\textrm{in}}+1)^{n+1}}|n\rangle\langle n|$ is a thermal state with the average number of photons $\bar{N}_{\textrm{in}}$ and $S(r)$ and $D(\alpha)$ are squeezing and displacement operators with squeezing coefficient $r$ and displacement $\alpha$ respectively. The average number of photons in such state, and thus also the cost, is given by $|\alpha|^{2}+(\bar{N}_{\textrm{in}}+\frac{1}{2})\cosh2r-\frac{1}{2}=\bar{n}$. A more convenient way to handle the above state is to use a phase space picture in which it can be specified by its first and second moments, i.e., by a displacement vector $\vec{x}_{\textrm{in}}=(\sqrt{2}\textrm{Re}\alpha,\,\sqrt{2}\textrm{Im}\alpha)$ and a covariance matrix
\begin{equation}
\sigma_{\textrm{in}}=\left(\bar{N}_{\textrm{in}}+\frac{1}{2}\right)\left(\begin{array}{cc}
\frac{1}{\omega_{\textrm{in}}} & 0\\
0 & \omega_{\textrm{in}}
\end{array}\right),\label{eq:cov_matrix}
\end{equation}
where I have assumed that off-diagonal terms are zero (it is always possible to find a basis in phase space in which such a form is right) and defined as $e^{-2r}=\omega_{\textrm{in}}$. A Gaussian channel acting on a Gaussian state transforms its first moments and covariance matrix $\vec{x}_{\textrm{out}}=\mathbf{X}\vec{x}_{\textrm{in}}+
\vec{x}_{\textrm{env}}$ and $\sigma_{\textrm{out}}=\mathbf{X}\sigma_{\textrm{in}}\mathbf{X}^T+\mathbf{Y}$. Here $\mathbf{X}$ and $\mathbf{Y}$ satisfy $\mathbf{Y}+\frac{i}{2}\left(\Omega-\mathbf{X}^T\Omega\mathbf{X}\right)\leq 0$, where $\Omega$ is the symplectic matrix, and together with $\vec{x}_{\textrm{env}}$ specify the channel. It can be proven that in terms of the capacity per channel use it is sufficient to consider a fiducial channel \cite{Schafer2013, Schafer2016}, whose action on the input Gaussian state results in a state whose displacement vector is given by $\vec{x}_{\textrm{out}}=\sqrt{|\eta|}(\sqrt{2}\textrm{Re}\alpha,\,\textrm{sgn}(\eta)\sqrt{2}\textrm{Im}\alpha)$ and covariance matrix
\begin{multline}\label{eq:cov_matrix_out}
\sigma_{\textrm{out}}=\left(\bar{N}_{\textrm{out}}+\frac{1}{2}\right)\left(\begin{array}{cc}
\frac{1}{\omega_{\textrm{out}}} & 0\\
0 & \omega_{\textrm{out}}
\end{array}\right)=\\=|\eta|\sigma_{\textrm{in}}+|1-\eta|\left(\tilde{N}+\frac{1}{2}\right)\left(\begin{array}{cc}
\frac{1}{\tilde{\omega}} & 0\\
0 & \tilde{\omega}
\end{array}\right),
\end{multline}
where intuitively $\eta=\det\mathbf X$ denotes the transmission (for $0\leq\eta\leq 1$) or gain ($\eta>1$ or $\eta<0)$ of the channel and $\bar{N}_{\textrm{out}}$ and  $\omega_{\textrm{out}}$ are average number of thermal photons and squeezing of the output state, respectively. The parameters $\tilde{N}$ and $\tilde{\omega}$ can be roughly thought of as the amount of thermal noise in the environment (this is strictly true for $\tilde{\omega}=1$) and squeezing of the environment, respectively. I will therefore consider only Gaussian channels of the above form.

The resulting relative entropy is given by eq.~(\ref{eq:entropy_gaussian_vacuum}) in the Appendix \ref{app:b} (see also \cite{Chen2005, Pirandola2017, Seshadreesan2017}). The capacity per unit cost is thus
\begin{equation}\label{eq:cap_gauss}
\mathbf{C}=\sup_{\bar{n}>0}\sup_{\bar{N}_{\textrm{in}},\omega_{\textrm{in}},\alpha}\frac{D(\rho_{\textrm{out}}||\rho_{0})}{\bar{n}}
\end{equation}
where the second supremum is taken over values $\bar{N}_{\textrm{in}},\,\omega_{\textrm{in}},\,\alpha$ satisfying the average energy constraint $|\alpha|^{2}+\frac{1}{2}(\bar{N}_{\textrm{in}}+\frac{1}{2})(\omega_{\textrm{in}}+\frac{1}{\omega_{\textrm{in}}})-\frac{1}{2}=\bar{n}$. Importantly, it can be shown that the supremum in the above formula is always attained in the limit $\bar{n}\to0$, so it gives also PIE in the weak signal power regime. Note that this means that in the weak power regime the ultimate PIE can be attained by just using generalized on-off keying modulation \cite{Kitayama2014}. Below, I will investigate some special cases.

\paragraph{Lossy channel}

If the evolution is just pure losses $\tilde{N}=0,\,\tilde{\omega}=1$ and $0\leq\eta\leq 1$, the free state does not change $\rho_{0}=|0\rangle\langle0|$ and remains pure. Thus, the relative entropy $D(\rho_{\textrm{out}}||\rho_{0})$ is infinite irrespective of the second input state since it must have different support than just vacuum state. This means that $\mathbf{C}_{\textrm{loss}}=\infty$ and PIE is unbounded. This is consistent with previous results \cite{Guha2011a, Jarzyna2015, Chung2016} and with the ordinary capacity of the lossy channel in particular \cite{Giovannetti2004a, Giovannetti2014}. In this case it is also easy to find a measurement scheme that saturates the bound. Since the second state has different support than the vacuum state it is sufficient to perform simple photodetection in order to obtain arbitrarily large PIE \cite{Guha2011a, Jarzyna2015}.

\paragraph{Phase insensitive channels}

\begin{figure}[t!]
\includegraphics[width=\columnwidth]{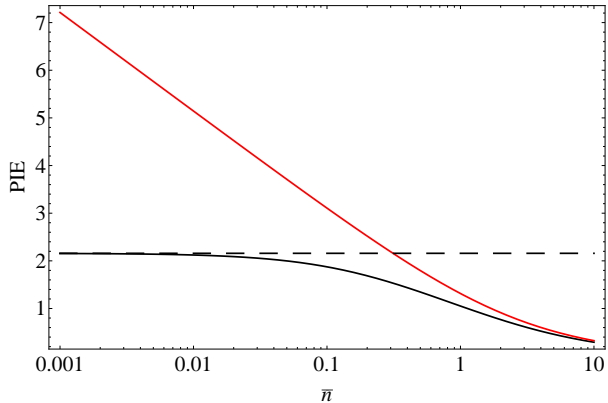}
\caption{Photon information efficiency as a function of the average number of photons $\bar{n}$ for a lossy (red) and thermal channel (black, solid) for $\eta=0.9$ and $\tilde{N}=1$. The dashed line indicates the capacity per unit cost for the thermal channel.}
\label{fig:figure2}
\end{figure}

For phase insensitive channels $\tilde{\omega}=1$, and, since there is no squeezing in the environment it is optimally to encode information in the displacement of coherent states \cite{Schafer2013}. This results in $\omega_{\textrm{out}}=\omega_{\textrm{out}}^0=1$ and $\bar{N}_{\textrm{out}}=\bar{N}_{\textrm{out}}^0=\frac{|\eta|-1}{2}+|1-\eta|(\tilde{N}+\frac{1}{2})$, where $\omega_{\textrm{out}},\,\bar{N}_{\textrm{out}}$ and $\omega_{\textrm{out}}^{0},\,\bar{N}_{\textrm{out}}^{0}$ are parameters of the output covariance matrix for the input in coherent and vacuum states respectively. The expression for relative entropy  eq.~(\ref{eq:entropy_gaussian_vacuum}) may be thus simplified to
\begin{equation}
D(\rho_{\textrm{out}}||\rho_{0})=|\eta||\alpha|^{2}\log\frac{\bar{N}_{\textrm{out}}^{0}+1}{\bar{N}_{\textrm{out}}^{0}}
\end{equation}
Since $|\alpha|^2=\bar{n}$, the capacity per unit cost is according to eq.~(\ref{eq:cap_gauss}) given by $\mathbf{C_{\textrm{ph}}}=|\eta|\log\frac{\bar{N}_{\textrm{out}}^{0}+1}{\bar{N}_{\textrm{out}}^{0}}$. Remarkably, this is also the result that one can obtain by using the formula $\mathbf{C}=\frac{1}{2}\mathcal{J}$ n eq.~(\ref{eq:cap_REQFI_eq}), where $\mathcal{J}$ is the REQFI for estimating the amplitude of the coherent state for the considered evolution. This means that unlike the lossy case, PIE converges to a finite maximum with weakening signal power. In fact, the lossy channel is the only Gaussian channel for which the capacity per unit cost and maximal PIE are infinite. This is caused by the fact that for other kinds of phase insensitive Gaussian evolutions the vacuum state encoding the free symbol is transformed into a thermal state supported on the whole Hilbert space. The PIE for two kinds of phase insensitive channels, a thermal and a lossy channel, is plotted in \figref{fig:figure2}. It can be easily seen that for a lossy channel PIE diverges to infinity for a decreasing average number of photons whereas for a thermal channel it saturates at the value given by the capacity per unit cost.

\paragraph{Squeezing channel}

The most general case is the squeezing channel $\tilde{\omega}\neq 1$. Similarly to the previous cases, the capacity per unit cost for the squeezing channel is attained by coherent states and reads $\mathbf{C_{\textrm{sq}}}=|\eta|\omega_{\max}\log\frac{\bar{N}_{\textrm{out}}^{0}+1}{\bar{N}_{\textrm{out}}^{0}}$, where $\omega_{\max}=\max\left(\omega_{\textrm{out}}^0,\,\frac{1}{\omega_{\textrm{out}}^0}\right)$ and the definitions of $\omega_{\textrm{out}}^0$ and $N_{\textrm{out}}^0$ are given in eq.~(\ref{eq:vacuum_coeffcients}). This indicates that PIE converges to a constant number with $\bar{n}\to0$. The presence of squeezing in the environment may be beneficial since it allows for larger capacities per unit cost, as $\mathbf{C}_{\textrm{sq}}\geq\mathbf{C_{\textrm{ph}}}$ for the same average number of thermal photons in both cases. Note also that for the squeezing channel coherent state encoding in general is not the best choice in terms of the capacity per channel use; it is optimal to use displaced squeezed states \cite{Schafer2016}. From the fact that it is possible to saturate the capacity per unit cost also with the coherent state ensemble it is evident that they can perform as well as displaced squeezed states in the regime of weak signal power.

\section{Conclusions}

In summary, I have shown that classical results on the capacity per unit cost can be easily generalized to describe the transmission of information encoded in quantum states and its relation to photon information efficiency. Specifically, for channels that allow for a cost-free input state I showed a simplified formula for the capacity per unit cost and argued that all channels that leave the free state pure have an infinite capacity per unit cost. Consequently, channels corrupting the free state state have a finite capacity per unit cost and bounded PIE indicating that an ordinary capacity per single use of the channel cannot scale better than linearly with the average number of photons per bin in the regime of weak signals. For a particular case of quadratic cost function I gave also a relation between the capacity per unit cost and quantities from quantum estimation theory, establishing a link between those two theories

After completing this work I became aware of a related work \cite{Ding2017} in which the authors generalize the concept of capacity per unit cost to the case of communication over quantum channels and additionally consider quantum communication in which one transmits quantum rather than classical information.

\section{Acknowledgments}

I thank Rafa{\l} Demkowicz-Dobrza{\'n}ski and Konrad Banaszek for many insightful discussions and comments on the manuscript. This work was supported by Polish Ministry of Science and Higher Education Iuventus Plus program for years 2015-2017 No. 0088/IP3/2015/73 as well as by the Foundation for Polish Science under the TEAM project "Quantum Optical Communication Systems" co-financed by the European Union under the European Regional Development Fund.

\bibliographystyle{apsrev4-1}

\appendix

\section{Explicit expression for relative entropy quantum Fisher information}\label{app:a}

Consider a family of states $\rho_\varphi$ parametrized by a real parameter $\varphi$. In order to find REQFI at $\varphi_0$ it is sufficient to find the second order Taylor expansion of the relative entropy $D(\rho_{\varphi_0+\delta\varphi}||\rho_{\varphi_0})=\frac{\delta\varphi^2}{2}\mathcal{J}\left[\rho_{\varphi_0}\right]+O(\delta\varphi^3)$. Such an expansion can be easily written by straightforward calculation in the eigenbasis of $\rho_{\varphi_0}=\sum_n p_n\ket{n}\bra{n}$ as
\begin{equation}
\mathcal{J}\left[\rho_{\varphi_0}\right]=\sum_n\frac{\dot{p}_n^2}{p_n}+2\sum_{n,k}(p_n-p_k)|\braket{n}{\dot{k}}|^2\log p_n,
\end{equation}
where overdots denote derivatives with respect to $\varphi$. Note, that if the support of the state derivative $\dot{\rho}_\varphi|_{\varphi=\varphi_0}$ has a part that lies outside the support of $\rho_{\varphi_0}$ then the second term of the above expression, and consequently REQFI, is infinite. Since $\dot{\rho}_\varphi\delta\varphi$ is roughly speaking the difference between states $\rho_{\varphi_0}$ and $\rho_{\varphi_0+\delta\varphi}$, this is in agreement with the fact that the relative entropy between two states with partially overlapping supports is infinite. 

\section{Relative entropy for Gaussian states}\label{app:b}

I will present here the way of obtaining relative entropy between two arbitrary one mode Gaussian states, which can be found also in \cite{Chen2005, Pirandola2017, Seshadreesan2017}. Given two density matrices $\rho_1$ and $\rho_2$, the relative entropy between them is equal to
\begin{equation}\label{eq:rel_entr}
D(\rho_1||\rho_2)=S(\rho_1)-\textrm{Tr}\rho_1\log\rho_2,
\end{equation}
where $S(\rho_1)$ is von Neumann entropy of state $\rho_1$. If the state is Gaussian than its entropy is equal to $S(\rho_1)=g(\gamma_1)$, where $\gamma_1$ is the symplectic eigenvalue of the covariance matrix $\sigma_1$ of the state and the function $g$ is $g(x)=(x+\frac{1}{2})\log(x+\frac{1}{2})-(x-\frac{1}{2})\log(x-\frac{1}{2})$ \cite{Adesso2006}. To evaluate the second term in eq.~(\ref{eq:rel_entr}) I will use the fact that
\begin{equation} 
\textrm{Tr}\rho_1\log U\rho_2 U^\dagger=\textrm{Tr}U^\dagger\rho_1 U\log\rho_2,
\end{equation}
which holds for any unitary transformation $U$. If the state $\rho_2$ has the displacement vector $\vec{r}_2$, then by taking $U=D(\vec{r}_2)$, where $D(\vec{r})$ denotes the operator of displacement by vector $\vec{r}$, one can obtain
\begin{equation}
\textrm{Tr}\rho_1\log\rho_2=\textrm{Tr}D^\dagger(\vec{r}_2)\rho_1 D(\vec{r}_2)\log\tilde{\rho}_2=\textrm{Tr}\tilde{\rho}_1\log\tilde{\rho}_2,
\end{equation}
where the state $\tilde{\rho}_2$ has the same covariance matrix as $\rho_2$ but vanishing first moments and $\tilde{\rho}_1$ is just $\rho_1$ with first moments shifted by $\vec{r}_2$. The Gaussian state with vanishing first moments can be easily expressed as
\begin{equation}
\rho=e^{-\frac{1}{2}\hat{\vec{R}}^T\mathbf{M}\hat{\vec{R}}}/Z,
\end{equation}
where $Z=\textrm{Tr}\left[e^{-\frac{1}{2}\hat{\vec{R}}^T\mathbf{M}\hat{\vec{R}}}\right]$ is the normalization constant, $\mathbf{M}$ is a symmetric positive matrix, and $\hat{\vec{R}}$ is a vector of quadrature operators. Using this expression for $\tilde{\rho}_2$, the second term in eq.~(\ref{eq:rel_entr}) reads
\begin{multline}
\textrm{Tr}\tilde{\rho}_1\log\tilde{\rho}_2=-\log Z-\frac{1}{2}\textrm{Tr}\tilde{\rho}_1\hat{\vec{R}}^T\mathbf{M}\hat{\vec{R}}=\\
=-\log Z-\frac{1}{2}\textrm{Tr}\hat{R}_i\tilde{\rho}_1\hat{R}_j\mathbf{M}_{ij}=\\
=-\log Z-\frac{1}{2}\textrm{Tr}\sigma_1\mathbf{M}-\frac{1}{2}\vec{x}^TM\vec{x},
\end{multline}
where $\vec{x}=\vec{r}_1-\vec{r}_2$ is the difference between first moments of states $\rho_1$ and $\rho_2$. The relative entropy is therefore given by
\begin{equation}\label{eq:relative_gaussian}
D(\rho_1||\rho_2)=-g(\gamma_1)+\log Z+\frac{1}{2}\textrm{Tr}\sigma_1 \mathbf{M}+\frac{1}{2}\vec{x}^T\mathbf{M}\vec{x}.
\end{equation}
The matrix $\mathbf{M}$ can be directly calculated from the covariance matrix of the state $\rho_2$. This is because $\mathbf{M}=(S^T)^{-1}\tilde{\mathbf{M}}S^{-1}$, where $S$ is the symplectic transformation diagonalizing $\sigma_2$ and $\tilde{\mathbf{M}}$ is a diagonal matrix with eigenvalues equal to $\log\frac{2\gamma_2+1}{2\gamma_2-1}$, where $\gamma_2$ is the symplectic eigenvalue of $\sigma_2$. Similarly, the normalization factor is given by $Z=\sqrt{\gamma_2^2-\frac{1}{4}}$.

Eq.~(\ref{eq:relative_gaussian}) may now be applied to calculate the relative entropy $D(\rho_{\textrm{out}}||\rho_{0})$ between states resulting from the evolution of some Gaussian state and vacuum state under the general Gaussian dissipative dynamics considered in the main text. The output state of the dynamics is given by covariance matrix in eq.~(\ref{eq:cov_matrix_out}) and the input by the one in eq.~(\ref{eq:cov_matrix}). To get the symplectic eigenvalue of the output covariance matrix it needs to be expressed in the form given by eq.~(\ref{eq:cov_matrix}). Straightforward algebra leads to
\begin{multline}
\omega_{\textrm{out}}=\sqrt{\frac{|\eta|(\bar{N}_{\textrm{in}}+\frac{1}{2})\frac{1}{\omega_0}+|1-\eta|(\tilde{N}+\frac{1}{2})\frac{1}{\tilde{\omega}}}{|\eta|(\bar{N}_{\textrm{in}}+\frac{1}{2})\omega_{\textrm{in}}+|1-\eta|(\tilde{N}+\frac{1}{2})\tilde{\omega}}},\\
\bar{N}_{\textrm{out}}+\frac{1}{2}=\qquad\qquad\qquad\qquad\qquad\qquad\qquad\qquad\\
=\sqrt{\left(|\eta|\left(\bar{N}_{\textrm{in}}+\frac{1}{2}\right)\frac{1}{\omega_{\textrm{in}}}+|1-\eta|\left(\tilde{N}+\frac{1}{2}\right)\frac{1}{\tilde{\omega}}\right)}\times\\
\times\sqrt{\left(|\eta|\left(\bar{N}_{\textrm{in}}+\frac{1}{2}\right)\omega_{\textrm{in}}+|1-\eta|\left(\tilde{N}+\frac{1}{2}\right)\tilde{\omega}\right)},
\end{multline}
where the second expression is the symplectic eigenvalue and I took the notation used in sec.~(\ref{sec:gaussian}). For the vacuum input state ($\bar{N}_{\textrm{in}}=0,\,\omega_{\textrm{in}}=1$) above equations simplify to
\begin{multline}\label{eq:vacuum_coeffcients}
\omega_{\textrm{out}}^0=\sqrt{\frac{\frac{|\eta|}{2}+\frac{|1-\eta|}{\tilde{\omega}}(\tilde{N}+\frac{1}{2})}{\frac{|\eta|}{2}+|1-\eta|(\tilde{N}+\frac{1}{2})\tilde{\omega}}},\\
\bar{N}_{\textrm{out}}^0+\frac{1}{2}=\sqrt{\left(\frac{|\eta|}{2}+\frac{|1-\eta|}{\tilde{\omega}}\left(\tilde{N}+\frac{1}{2}\right)\right)}\times\\
\times\sqrt{\left(\frac{|\eta|}{2}+|1-\eta|\left(\tilde{N}+\frac{1}{2}\right)\tilde{\omega}\right)}.
\end{multline}
Using these expressions, the matrix $\mathbf{M}$ for the vacuum state is given by
\begin{equation}
\mathbf{M}=\left(\begin{array}{cc}
\omega_{\textrm{out}}^{0} & 0\\
0 & \frac{1}{\omega_{\textrm{out}}^{0}}
\end{array}\right)\log\frac{\bar{N}_{\textrm{out}}^{0}+1}{\bar{N}_{\textrm{out}}^{0}}.
\end{equation}
Denoting by $\alpha$ the displacement of the input Gaussian state, the final expression for relative entropy between the arbitrary Gaussian state and vacuum state subjected to general Gaussian dissipative dynamics reads
\begin{multline}\label{eq:entropy_gaussian_vacuum}
D(\rho_{\textrm{out}}||\rho_{0})=\bar{N}_{\textrm{out}}\log\bar{N}_{\textrm{out}}+\\
-(\bar{N}_{\textrm{out}}+1)\log(\bar{N}_{\textrm{out}}+1)+\frac{1}{2}\log\bar{N}_{\textrm{out}}^{0}(\bar{N}_{\textrm{out}}^{0}+1)+\\
+\left(\bar{N}_{\textrm{out}}+\frac{1}{2}\right)\frac{(\omega_{\textrm{out}}^{0})^2+\omega_{\textrm{out}}^2}{2\omega_{\textrm{out}}^{0}\omega_{\textrm{out}}}\log\frac{\bar{N}_{\textrm{out}}^{0}+1}{\bar{N}_{\textrm{out}}^{0}}+\\
+|\eta|\left((\textrm{Re}\alpha)^{2}\omega_{\textrm{out}}^{0}+\frac{(\textrm{Im}\alpha)^{2}}{\omega_{\textrm{out}}^{0}}\right)\log\frac{\bar{N}_{\textrm{out}}^{0}+1}{\bar{N}_{\textrm{out}}^{0}}.
\end{multline}

\end{document}